\documentclass[12pt,preprint]{aastex}

\makeatletter

\newcommand{\Rmnum}[1]{\expandafter\@slowromancap\romannumeral#1@}
\makeatother
\usepackage{amsmath,amssymb}

\slugcomment{Accepted for publication in ApJ}
\shorttitle{Simulating solar prominences}
\shortauthors{Xia et al.}

\begin{document}
\title{Formation and plasma circulation of solar prominences}
\author{C. Xia\altaffilmark{1}, R. Keppens\altaffilmark{1,2}}

\altaffiltext{1}{Centre for mathematical Plasma Astrophysics, Department of
Mathematics, KU Leuven, Celestijnenlaan 200B, 3001 Leuven, Belgium}
\altaffiltext{2}{School of Astronomy and Space Science, Nanjing University, Nanjing 210093, China}

\begin{abstract}
Solar prominences are long-lived cool and dense plasma curtains in the hot and 
rarefied outer solar atmosphere or corona. The physical mechanism responsible 
for their formation and especially for their internal plasma circulation
has been uncertain for decades. The observed 
ubiquitous down flows in quiescent prominences are difficult to interpret as 
plasma with high conductivity seems to move across horizontal magnetic field lines.
Here we present three-dimensional numerical simulations of prominence formation 
and evolution in an elongated magnetic flux rope as a result of in-situ plasma
condensations fueled by continuous plasma evaporation from the solar chromosphere. The
prominence is born and maintained in a fragmented, highly dynamic state with 
continuous reappearance of multiple blobs and thread structures that move mainly 
downward dragging along mass-loaded field lines. The prominence plasma circulation is 
characterized by the dynamic balance between the drainage of prominence plasma 
back to the chromosphere and the formation of prominence plasma via continuous 
condensation. Plasma evaporates from the chromosphere, condenses into the 
prominence in the corona, and drains back to the chromosphere, establishing a stable 
chromosphere-corona plasma cycle. Synthetic images of the modeled prominence with 
the Solar Dynamics Observatory Atmospheric Imaging Assembly closely resemble 
actual observations, with many dynamical threads underlying an elliptical 
coronal cavity.
\end{abstract}


\section{INTRODUCTION}\label{intro}
Solar prominences are about 100 times denser and cooler than their surrounding hot
corona, with their weight supported by the magnetic field against gravity. They 
exist for days to weeks above polarity inversion lines (PIL), which 
separate positive and negative magnetic flux regions in the photosphere. 
Prominences are frequently found inside coronal cavities, which are tunnel-like 
elliptical dark regions \citep{Gibson10} at the solar limb, with up to 40 \% density 
depletion \citep{Fuller08}, around the prominences. The magnetic 
structure hosting a prominence and its cavity is believed to be a helical magnetic 
flux rope confined by overlying magnetic loops, which is consistent with observed 
spinning plasma motions \citep{WangYM10} as well as linear polarization 
signatures \citep{Bak13}. A flux rope was present in early two-dimensional (2D) theoretical 
prominence models \citep{Kupe74} and 2D magnetohydrostatic solutions for 
prominence-embedded flux ropes have been demonstrated analytically \citep{Low04} 
and numerically \citep{Blok11,Hillier13}.

Static prominence models assume that quiescent prominences change little with 
time. However, high resolution observations 
have found that quiescent prominences consist of fine and vertically oriented 
plasma threads, that continuously evolve with downward motions (4 to 35 km 
s$^{-1}$) \citep{Engvold76,Berger08,Chae08}. Moreover, episodic dark plumes may 
rise \citep{Berger08,Berger11} from the bottom of prominences and propagate 
upward between threads with speeds about 10 to 20 km s$^{-1}$. The heavy 
prominence material suspended above the light coronal plasma can be liable to 
Rayleigh--Taylor (RT) instability \citep{Ryutova10} which has been investigated 
in several numerical magnetohydrodynamic (MHD) models \citep{Khomenko14,
Hillier11,Terradas15,Keppens15}. However, none of these models addressed 
prominence formation and mass cycling, or studied a realistic 3D flux rope. 

The idea that prominences form via in situ condensation of coronal plasma was 
inferred from thermal instability theory \citep{Parker53,Field65} and is supported 
by observations in extreme ultraviolet (EUV) channels \citep{Liu12,Berger12}.  
Moreover, continuous condensations must happen to compensate mass loss of 
quiescent prominences, which requires an incessant supply of hot plasma from 
chromosphere to corona. Such hot plasma flows have not been convincingly 
detected, although persistent horizontal flows of H$\alpha$-emitting cool plasma
originating from the chromosphere into a prominence was observed \citep{Chae08} 
and the episodic dark rising plumes filled with hot plasma may provide a small
fraction of the needed mass \citep{Berger10}.
Numerical modeling of condensations as a result of runaway radiative cooling 
started with one-dimensional hydrodynamic simulations along individual 
magnetic field lines \citep{antiochos99,Karpen01,Xia11,Luna12} and progressed in 
recent years to 2D MHD simulations in magnetic arcades \citep{Xia12,Keppens14}.
These models add localized heating at footpoints of magnetic loops to 
evaporate plasma from chromosphere to corona leading to continuous 
condensations. However, restricted by 1D or 2D setups, they cannot fully establish 
the internal dynamics of prominences. Recent work on prominence formation via
in situ condensation in a 2D \citep{Kaneko15} and 3D 
flux rope \citep{Xia14} did not reproduce internal dynamics and mass drainage 
and only found a small prominence fragment that settled in a magnetostatic 
equilibrium.

We here report a set of 3D models that for the first time ever demonstrate
continuous formation of prominence condensations in a coronal cavity, 
achieving a balanced chromosphere-corona plasma cycle characterized by
vertical flows in thin prominence threads. The numerical methods and the simulation 
procedure are explained in Section~\ref{meth}. We present the results in 
Section~\ref{resu}. Finally, the results are discussed in Section~\ref{conc}.

\section{SIMULATION STRATEGY}\label{meth}
Since the magnetic topology of quiescent prominences is believed to be an elongated
magnetic flux rope, we first perform isothermal MHD simulations to simulate the 
formation of a magnetic flux rope driven by footpoint motions and flux cancellation 
 \citep{vanBalleg89,Xia14a}.
Our 3D model is set up in a Cartesian box covering 200 Mm $\times$ 120 Mm $\times$ 80 Mm 
, in $x$ (-100 Mm to 100 Mm), $y$ (-60 Mm to 60 Mm), and $z$ (0 to 80 Mm) directions, 
where a finite beta corona of 1 MK constant temperature is initially constructed with 
density and gas pressure prescribed from hydrostatic equilibrium with a bottom 
$10^9$ cm$^{-3}$ number density. We start from a bipolar magnetic 
field in the shape of an elongated sheared arcade. This arcade field itself is 
generated by linear force-free field extrapolation from an analytically 
prescribed bipolar magnetogram with two opposite polarity regions shaped like 
two parallel baguettes. This magnetogram $B_m$ is formulated as:
\begin{equation*}
B_m=\begin{cases} B_0\exp(-\frac{(x-x_1+x_L)^2}{2\delta x^2}-\frac{(y-y_1)^2}{2\delta 
y^2})-B_0\exp(-\frac{(x-x_2+x_L)^2}{2\delta x^2}-\frac{(y-y_2)^2}{2\delta y^2}) &
\text{if $x<x_1-x_L$} \\
B_0\exp(-\frac{(y-y_1)^2}{2\delta y^2})-B_0\exp(-\frac{(y-y_2)^2}{2\delta y^2}) &
\text{if $x_1-x_L \le x \le x_1+x_L$} \\
B_0\exp(-\frac{(x-x_1-x_L)^2}{2\delta x^2}-\frac{(y-y_1)^2}{2\delta 
y^2})-B_0\exp(-\frac{(x-x_2-x_L)^2}{2\delta x^2}-\frac{(y-y_2)^2}{2\delta y^2}) &
\text{if $x>x_1+x_L$} \\
\end{cases}
\end{equation*}
where $B_0=20$ G, $x_1=x_2=0$ Mm, $y_1=-y_2=25$ Mm, $x_L=50$ Mm, $\delta x=10$ Mm, 
and $\delta y=13$ Mm. An exact Green's function method \citep{Chiu77} is used to
generate an initial linear force-free field with constant $\alpha=-0.08$. Since 
our simulation box is intended to start at low chromospheric heights, this 
magnetogram is placed 4 Mm below the bottom plane of the simulation box.

A composite surface flow is imposed at the bottom boundary with the formula as:
\begin{equation*}
 v_y^b=-f(t) C_1 \frac{\partial |B_m|}{\partial y} \exp(-y^2/y_d^2),
 v_x^b=-v_y^b
\end{equation*}
where $y_d = 35$ Mm quantifies an additional Gaussian width parameter away from 
the polarity inversion line (PIL), $f(t)$ is a linear ramp function to switch 
on and off the driving flow, and 
the amplitude factor $C_1$ is chosen so that the driving speed 
has a maximum value of 12.8 km s$^{-1}$ and the maximum initial Alfv\'{e}n Mach 
number is 0.0155. The flow is a combination of a shearing flow and a converging 
flow relative to the PIL. The shearing flow roughly mimics the effective 
shearing of an east-west bipolar arcade by the differential rotation of the solar 
surface. The converging flow from the strong field regions to weak field regions
 is an effective result of magnetic element diffusion caused by random 
supergranular motions \citep{Leighton64}. Flows far away from the PIL are 
suppressed by the exponential factor for an easier handling of the side boundaries.

The model evolution is performed by solving the isothermal MHD equations given by
\begin{align}
 \frac{\partial \rho}{\partial t}+\nabla\cdot\left(\rho\mathbf{v}\right)
       &=0,\\
 \frac{\partial \left(\rho\mathbf{v}\right)}{\partial t}+\nabla\cdot\left(
  \rho\mathbf{vv}+p_{\rm tot}\mathbf{I}-\frac{\mathbf{BB}}{\mu_0}\right)&=
  \rho\mathbf{g},\\
 \frac{\partial \mathbf{B}}{\partial t}+\nabla\cdot\left(\mathbf{vB}-
  \mathbf{Bv}\right)&=0,
\end{align}
where $\rho$, $\mathbf{v}$, $\mathbf{B}$, and $\mathbf{I}$ are the plasma
density, velocity, magnetic field, and unit tensor, respectively, while the
total pressure is $p_{\rm tot}\equiv p+\frac{B^2}{2 \mu_0}$ and $\mathbf{g}=-
g_\odot r_\odot^2/(r_\odot+z)^2\mathbf{\hat{z}}$ is the gravitational
acceleration with $r_\odot$ the solar radius and $g_\odot$ the solar surface
gravitational acceleration.
To normalize the equations for computation, we use 10 Mm, $10^9$ cm$^{-3}$, 
116.45 km s$^{-1}$ , and 2 Gauss as the unit of length, number density, velocity, 
and magnetic field, respectively. We use the Adaptive Mesh Refinement (AMR) 
Versatile Advection Code (MPI-AMRVAC) \citep{amrvac12,Porth14} to
numerically solve these equations with a third-order accurate scheme combining a 
Harten-Lax-van Leer scheme \citep{Harten83} with a third-order limited 
reconstruction \citep{Cada09} and a three-step Runge--Kutta time integration. 
To numerically maintain close to zero divergence of the magnetic field, we 
use the Generalized Lagrangian Multiplier method \citep{Dedner02}. The model 
domain at this phase is discretized on a three-level AMR grid which has an 
effective resolution of $400\times240\times240$ with the smallest cells
of 500 km $\times$ 500 km $\times$ 333 km. The simulation setup is completed by
the following boundary conditions. We force zero vertical velocity on the
bottom face and zero velocity on the other five faces of the box by 
antisymmetric boundary conditions. We extrapolate the magnetic field at
the bottom boundary and the top boundary from inner physical values, keeping the 
normal gradient zero, with a one-sided third-order finite difference representation. 
Then we modify the normal component to enforce the divergence-free condition 
in a second-order centered difference evaluation. The magnetic fields at the 
four side boundaries are kept fixed. The density values in the side boundary 
cells are copied from the neighbouring cells of the physical domain.
We fix the gravitationally stratified density at the bottom and adopt a 
gravitationally stratified density profile at the top. 

\begin{figure}
\includegraphics[width=\textwidth]{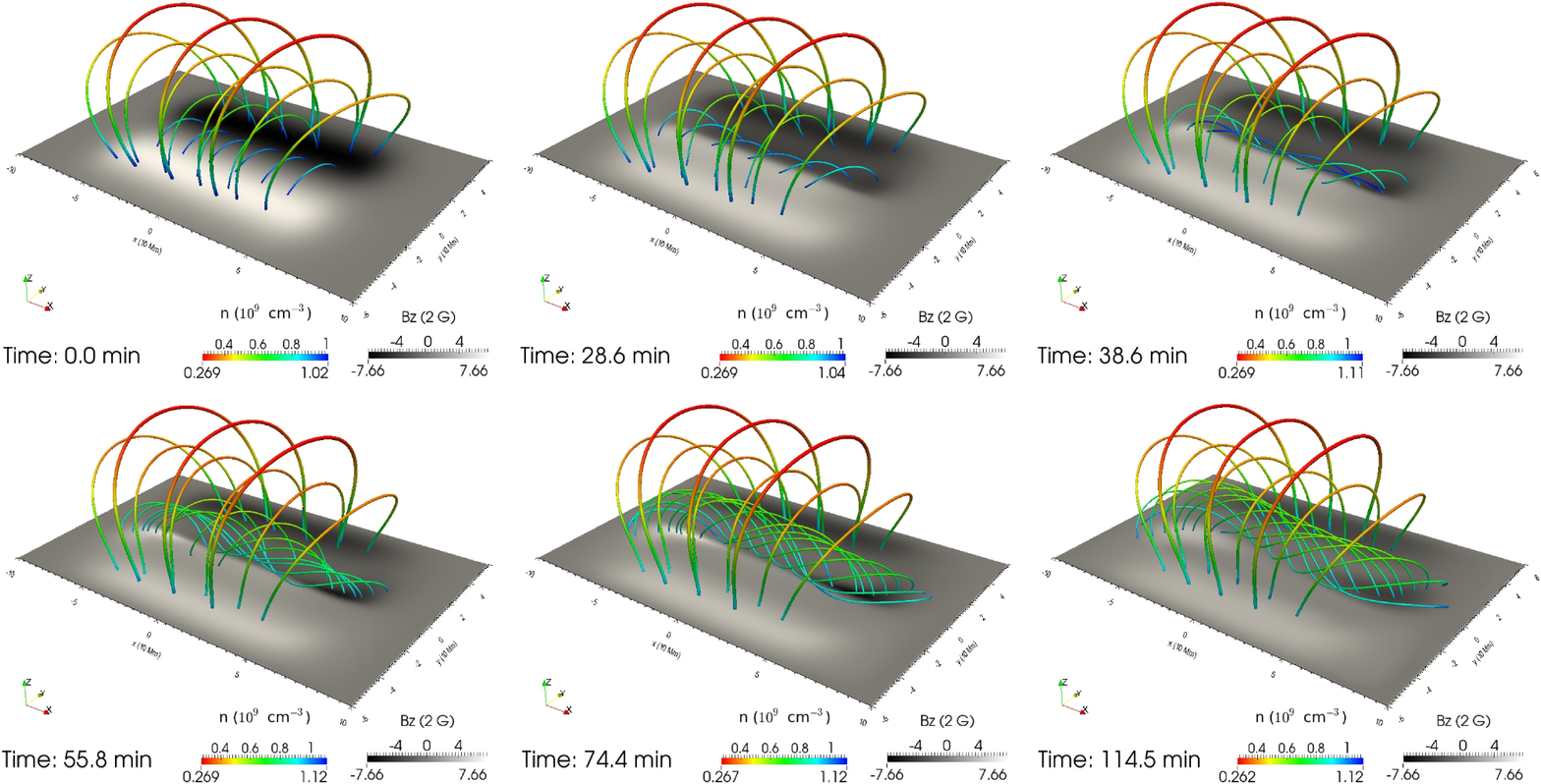}
\caption{The formation process of a magnetic flux rope in the solar corona.
In these panels depicting six subsequent moments, magnetic field lines colored by number density ($n$) show the forming flux 
rope and overlying arcade loops and greyscale of the bottom plane indicating vertical magnetic field strength.
}
\label{ffrform}
\end{figure}

The bottom flows drive those coronal loops that root in the inner halves of the 
main arcade polarities to move towards aligning with the PIL. As footpoints from 
opposite polarities collide on the PIL, magnetic reconnections occur due to 
finite numerical resistivity, to join arched loops into helical loops in a 
head-to-tail style. These helical loops, winding about the same central axial 
field line, form a helical magnetic flux rope (see Figure~\ref{ffrform}). We 
completely switch off the driving flows after 100 minutes and let the flux 
rope relax to an equilibrium state. 
In the end of this stage, a large-scale elongated flux rope is formed with 
its flux surface touching the bottom plane. The length of the flux rope is 
about 160 Mm and the diameter of its cross section is about 34 Mm. The density 
distribution does not change much compared to the initial state and 
the magnetic field has negligible Lorentz force.The magnetic field strength of 
the flux rope is about 6 to 7 Gauss which is consistent with the field strength
of a quiescent prominence since spectral polarimetric measurements show that 
the magnetic field in a quiescent prominence ranges from 3 to 30 G and it is 
predominantly horizontal making an acute angle with respect to the axis of 
the prominence \citep{Leroy83,Bommier94,Orozco14}.

In the second stage, we perform a full MHD simulation to produce a modeled 
solar atmosphere in thermal equilibrium containing the magnetic flux rope which was 
generated in the isothermal MHD modeling described above. To setup the initial 
state based on the isothermal flux rope, we kept the magnetic field unchanged 
and modified the density and temperature profiles from their isothermal states 
to a static vertically stratified solar atmosphere including chromosphere, 
transition region, and corona. We set the region below a height of 2.7 Mm to 
have chromospheric temperature 9600 K and above this height, the temperature 
increases with height in such a way that the vertical thermal conduction flux 
has a constant value of $2\times10^5$ erg cm$^{-2}$ s$^{-1}$. The density is 
then derived assuming a hydrostatic atmosphere with the number density at the 
bottom being $1.1\times10^{13}$ cm$^{-3}$. We now consider the full MHD 
equations with the energy equation as:
\begin{equation}
\frac{\partial E}{\partial t}+\nabla\cdot\left(E\mathbf{v}+
     p_{\rm tot}\mathbf{v}-\frac{\mathbf{BB}}{\mu_0}\cdot\mathbf{v}\right)=\rho\mathbf{g}\cdot
     \mathbf{v}+H-R+\nabla\cdot\left(\boldsymbol{\kappa}\cdot\nabla T\right), \\
\end{equation}
where total pressure $p_{\rm tot}\equiv p+B^2/2\mu_0$, gas pressure $p=2.3 n_{\rm  H} 
k_{\rm B} T$, and total energy $E=p/(\gamma-1)+\rho v^2/2+B^2/2\mu_0$. We again
use MPI-AMRVAC with the same combination of schemes as the isothermal model. The 
field-aligned ($\boldsymbol{\kappa}=\kappa_\parallel \mathbf{\hat{b}\hat{b}}$) 
thermal conduction is solved separately using a Runge-Kutta type Super 
Time Stepping scheme \citep{Meyer12} and $\kappa_\parallel=10^{-6}T^{5/2}$ erg 
cm$^{-1}$ s$^{-1}$ K$^{-1}$ is the Spitzer conductivity. Optically thin radiative 
cooling $R=1.2 n_{\rm H}^2\Lambda(T)$ is added using an exact integration 
scheme \citep{Townsend09}. To maintain a hot corona, the coronal heating is simulated by 
adding the parametrized heating term $H=H_0=c_0 e^{-z/\lambda}$ with 
$c_0=10^{-4}$ erg cm$^{-3}$ s$^{-1}$ and $\lambda=60$ Mm. We increased the highest 
AMR level of the mesh to 4, resulting in an effective resolution of 
$800\times480\times480$ with the smallest cells of 250 km $\times$ 250 km 
$\times$ 166 km. We modified the boundary conditions from the isothermal model to 
set zero velocity and fixed magnetic field at all boundaries. The gas pressure is 
fixed at the bottom and flexible at the top according to gravitational 
stratification. At side boundaries, the gas pressure copies values from the inner 
layer in the physical domain. This modified plasma state is not in thermal 
equilibrium, so we numerically solve the full MHD equations involving 
optically thin radiative loss, anisotropic thermal conduction and coronal heating 
until the system relaxes to a quasi-equilibrium after 114.5 minutes. 
Our endstate in this phase is reached when all thermal quantities settle into 
a stable 3D configuration where the magnitude of the remaining velocity is less 
than 10 km s$^{-1}$ and it is representative of a stable flux rope configuration 
with hot coronal plasma trapped inside it, where twisted field lines connect 
chromospheric to coronal plasma regions. 

In the third stage, we aim to form the macroscopic prominence in the stable 
flux rope obtained in the second stage. In order to simulate the chromospheric 
evaporation, which is an efficient way to bring material from the chromosphere to 
the corona, a relatively strong localized heating $H_1$ is added to the energy 
equation in addition to the global coronal heating $H_0$, namely, $H=H_0+H_1$. 
We only add the localized heating $H_1$ in two fixed cylinders which coincide with 
the two footpoint regions of the flux rope. These two cylinders are centered on 
$(x,y)$ positions on the bottom plane at (64, 6) Mm and (-64, -6) Mm, with a radius 
of 20 Mm. Within them, $H_1$ is constant in the chromosphere and decays rapidly 
with height above the approximate location of the transition region: 
\begin{equation*}
H_1=\begin{cases}f(t) c_1 e^{-((z-zh)/H_m)^2}  & \text{if $z > zh$} \\
      f(t) c_1  & \text{if $z \le zh$}
    \end{cases}
\end{equation*}
where $c_1=10^{-2}$ erg cm$^{-3}$ s$^{-1}$, $zh=5$ Mm and $H_m=3.16$ Mm. This 
final stage is simulated for a total of over four hours, and it is this stage 
which is presented in what follows.

\section{RESULTS}\label{resu}

\begin{figure}
\includegraphics[width=5.8in]{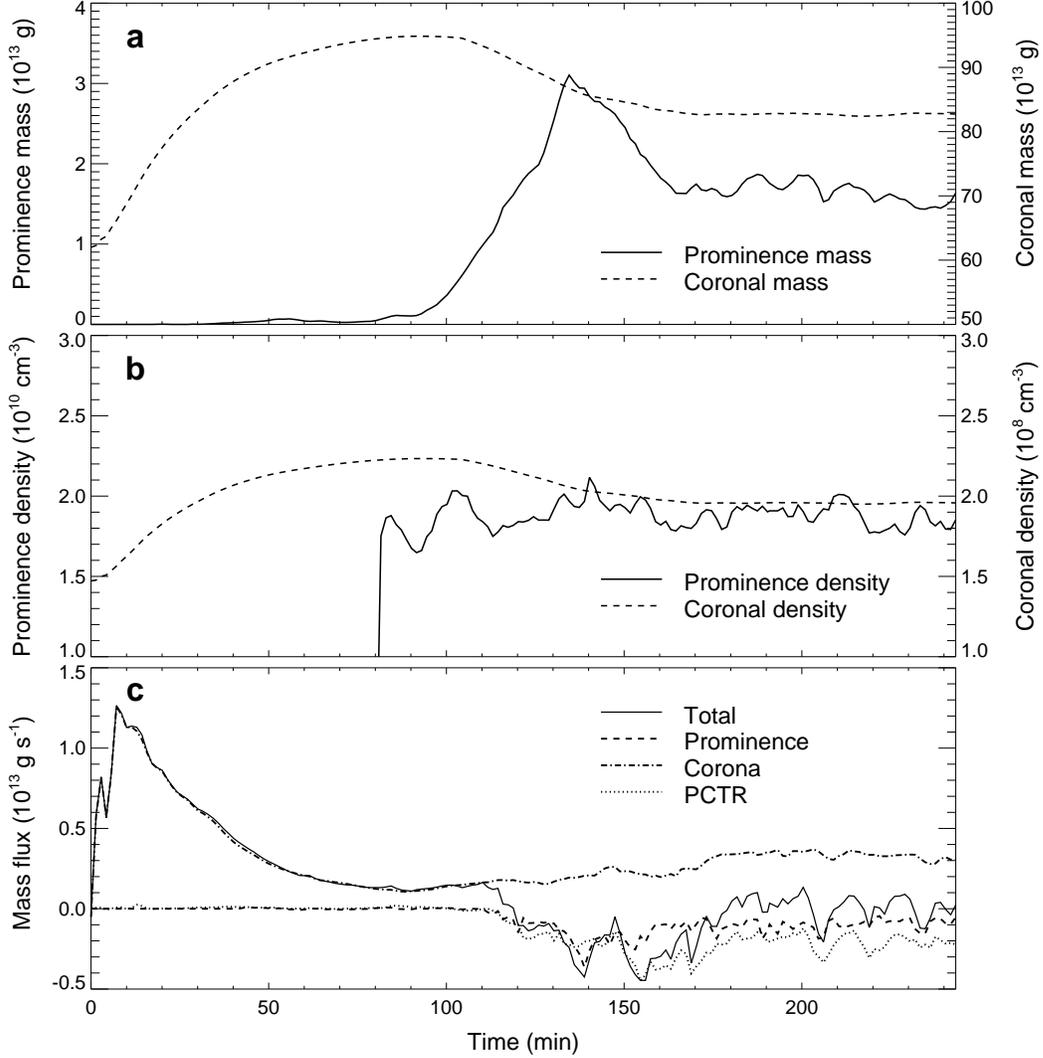}
\caption{Quantifying the formation and mass circulation of a virtual 
prominence. 
\textbf{a}, evolution of prominence mass (solid line) and coronal mass (dashed 
line) in the box. \textbf{b}, evolution of average plasma number density of 
the prominence (solid line) and the corona (dashed line). \textbf{c}, evolution 
of mass flux, through the bottom coronal plane ($z=8$ Mm), of prominence
plasma (dashed line), coronal plasma (dashed dotted line), prominence-corona
transition region (PCTR) (dotted line), 
and all plasma together (solid line). The PCTR is where the plasma density is 
lower than $10^9$ cm$^{-3}$ and temperature lower than 100,000 K.
}
\label{fmass}
\end{figure}

As chromospheric plasma 
gets evaporated into the corona, the lower part of the flux rope slowly evolves 
into a thermally unstable situation due to the dominating radiative cooling.
After about 100 minutes of gradual evolution, we then witness the runaway 
cooling stage when the decrease of temperature amplifies the radiation, leading
to the formation of dynamic blobs and threads. 
To get an overall impression of the evolution, we quantify the temporal evolution of
the mass, the average density, as well as the average velocity of all prominence 
plasma and all coronal plasma in the computational box (see Figure~\ref{fmass}a,b). In this 
quantification, prominence plasma has its number density higher than $10^{10}$ cm$^{-3}$,
its temperature lower than 20,000 K, and its altitude higher than 8 Mm. The 
change in local density and temperature conditions ultimately reaches the threshold to 
trigger thermal instability, with plasma condensations forming at a rapid pace 
after 80 minutes, accompanied by a decrease in coronal mass. At time 134.5 minutes, 
the mass of the prominence reaches a maximal $3.1\times10^{13}$ g, to then start 
decreasing as some blobs have by now fallen into the chromosphere. After 164.5 
minutes, the mass of the prominence settles and starts to oscillate around an 
average value of $1.65\times10^{13}$ g while the mass of the coronal volume reaches 
a stable value of $8.3\times10^{14}$ g. The number density of the prominence 
instantly adopts a value consistent with observations, as it oscillates around a 
mean value of $1.88\times10^{10}$ cm$^{-3}$. The average temperature of the 
prominence and the corona are 17,500 K and 1,660,000 K, respectively. For the 
entire 80 minutes following, we see lots of dynamics in the prominence as a whole, 
but the drainage of prominence mass into the chromosphere is balanced by new-born 
prominence material in continuously forming condensations. 

If we concentrate on the later stable phase after 164.5 minutes, the average 
vertical velocity of the prominence plasma is $-4.87$ km s$^{-1}$, showing that 
the prominence plasma is generally descending at a speed much slower than free-fall 
speed, which is consistent with observations \citep{Liu12}. To understand the mass flows, we quantify the mass flux through a horizontal 
plane at the bottom of the corona, including the total mass flux as well as 
contributions from prominence plasma, coronal plasma, and prominence-corona 
transition region (PCTR) plasma (Figure~\ref{fmass}c). Because of the chromospheric 
evaporation, the upward mass flux increases in the first 7 minutes and then 
decreases, impeded by the build-up of a high gas pressure inside the flux rope. 
As soon as the condensations begin, the evaporating mass flow 
increases slowly and then gradually settles at around 
$3.1\times10^{12}$ g s$^{-1}$. Prominence plasma with surrounding PCTR plasma starts 
to fall down through this plane back into the chromosphere after 110 minutes. The 
corresponding mass fluxes of prominence plasma and PCTR plasma oscillate with time and eventually stabilize at around 
$-10^{12}$ g s$^{-1}$ and $-2.1\times10^{12}$ g s$^{-1}$, respectively.
The total mass flux oscillates around zero, implying a dynamical 
equilibrium where a steady mass cycling is fully operational. We can draw the 
analogy with the water cycle on Earth, since we have solar chromospheric plasma that 
evaporates from the footpoints of the flux rope to reach higher altitudes into the 
corona where radiative condensations accumulate mass in prominence blobs that 
ultimately fall back to the chromosphere.

\begin{figure}
\includegraphics[width=\textwidth]{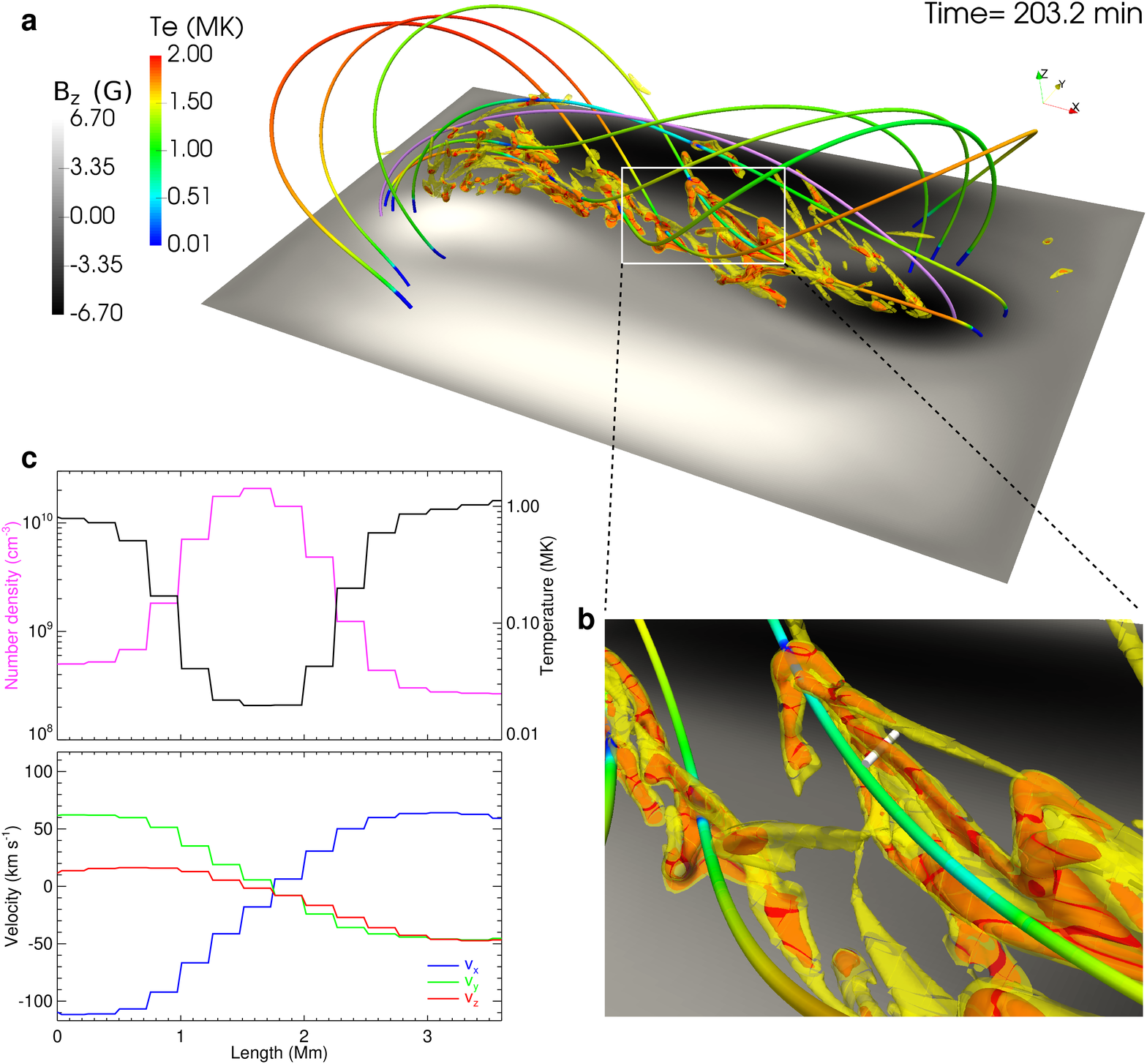}
\caption{Prominence,  magnetic field lines and shearing flows at time 
203.2 minutes. \textbf{a}, a global view of the prominence with the yellow 
translucent density contours at $4\times10^{9}$ cm$^{-3}$, the red density 
contours at $10^{10}$ cm$^{-3}$, magnetic field lines (colored by temperature 
Te) threading through the prominence, a purple magnetic field line representing 
the axis of the flux rope, and the bottom plane colored by $B_z$. \textbf{b}, 
close-up view of the region in white rectangular box in \textbf{a} with a white 
straight line cutting through a thread and the number density (pink curve), 
temperature (black curve) and velocities (blue curve $V_x$, green curve $V_y$, 
red curve $V_z$) along the cutting line displayed in \textbf{c}. 
}
\label{ft142}
\end{figure}

The most striking finding of our model is the extremely dynamic and intrinsically 
fragmented appearance of the prominence matter, from its first instant of formation 
throughout its steady plasma cycle. The prominence plasma is born in thin threads and 
blobs distributed over different places in the flux rope as visualized for a 
particular instant in Figure~\ref{ft142}a. Some 
of the blobs appear and then reside in the central dipped region of fairly 
left-right symmetrical, twisted field lines. Some blobs form and then surf along 
the shallow elbows of asymmetrically twisted field lines. A few blobs also pop up 
and then seemingly follow arched regions of weakly twisted field lines. Although 
every condensation has a different size and shape, the cross sections perpendicular 
to their long axis are roughly round. The diameter of these cross sections range 
from 1000 km to 1800 km. A typical variation of the temperature, density and 
velocity field across an individual blob is shown for the thread in the zoomed 
panel of Figure~\ref{ft142}b. There are typically shearing flows at both sides of the 
thread (Figure~\ref{ft142}c). These flows themselves influence the thread dynamical 
evolution and the velocity difference across the thread reaches up to 115 km 
s$^{-1}$ with local Alfv\'{e}n Mach number up to 1.

\begin{figure}
\includegraphics[width=\textwidth]{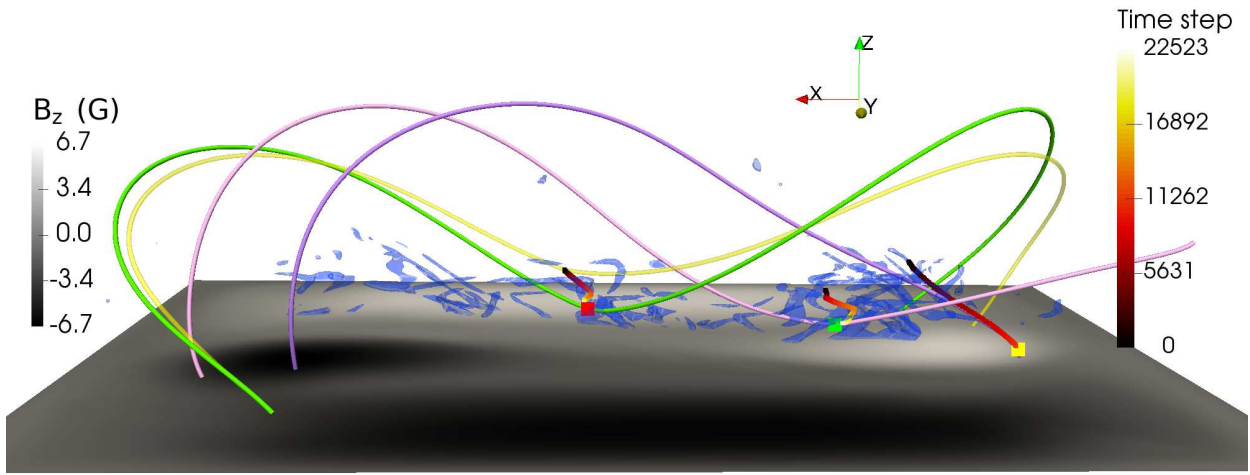}
\caption{Falling prominence blobs dragging down magnetic field lines. 
At time 143.1 minutes, three particle tracers, shown as the red, green, and 
yellow small squares, are added in prominence blobs (translucent blue contours). 
At a later time 214.7 minutes, these particle tracers descend with trajectories 
colored by the time step of particles. The red and green traced blobs drag down 
the green and pink field lines in their dips while the yellow marked blob slides 
along the purple field line. The translucent yellow field line is threading 
through the red tracer at time 143.1 minutes. 
} 
\label{ftracer}
\end{figure}

We already mentioned that the average vertical velocity in prominence matter is 
a few kilometers per second downwards. To see how the complex plasma motions of 
prominence fragments interact with the 
magnetic field, we use particle tracers to follow fluid elements. At time
143.1 minutes, we position massless and uncharged particles, which move passively 
with the local velocity field, in the cores of prominence threads, where plasma 
density is higher than $2\times10^{10}$ cm$^{-3}$ with temperature lower than 
20,000 K and plasma $\beta$ (the ratio between gas and magnetic pressure) ranges 
from 0.2 to 0.25. Each particle evolution involves integrating the advection 
equation with velocity information interpolated from the MHD run. In this 
integration process we introduce particle time steps different from
fluid time steps (limited by Courant--Friedrichs--Lewy condition) adopted in the time integration 
of the MHD equations: within one MHD fluid time step, the particles evolve more 
than one particle time step since particles are restricted to travel a distance 
shorter than the size of a grid cell in one particle time step. To evaluate the 
velocity of a particle at a particle time between two fluid time steps, we need 
both spatial and temporal interpolations on the velocity data from the MHD run. 
We first get two velocities at the particle position by linear spatial 
interpolation of the fluid velocity, one for each fluid time step. Then we do 
another linear interpolation in time between these two velocities to quantify 
the instantaneous local velocity field. We solve the equation of motion of the 
particle itself using an adaptive step size fourth-order Runge--Kutta scheme.
As the magnetic field is frozen-in the plasma due to the high conductivity in the 
solar atmosphere, these tracer particles allow us to disentangle the movement of 
individual magnetic field lines and the fluid dynamics. We find that generally,
those condensations that form and collect in the dips of helical magnetic field 
lines move downward under gravity dragging the field lines down to make the dips 
deeper (Figure~\ref{ftracer}). Note that the translucent yellow field line at 
time 143.1 minutes is bent by the falling blob to become the green field line at a 
later time 214.7 minutes. A typical descending speed is about 6 km s$^{-1}$ 
for the central condensations. On the other hand, condensations forming in the 
legs of arched magnetic field lines slip down along the field lines without 
significant deformation of these field lines. The heated arcade loops of the 
flux rope rise against the overlying constraining magnetic arcade, and this also 
leads to interchange motions with upward rising flux rope loops and 
downward sinking coronal loops in higher up regions above the prominence.

\begin{figure}
\includegraphics[width=\textwidth]{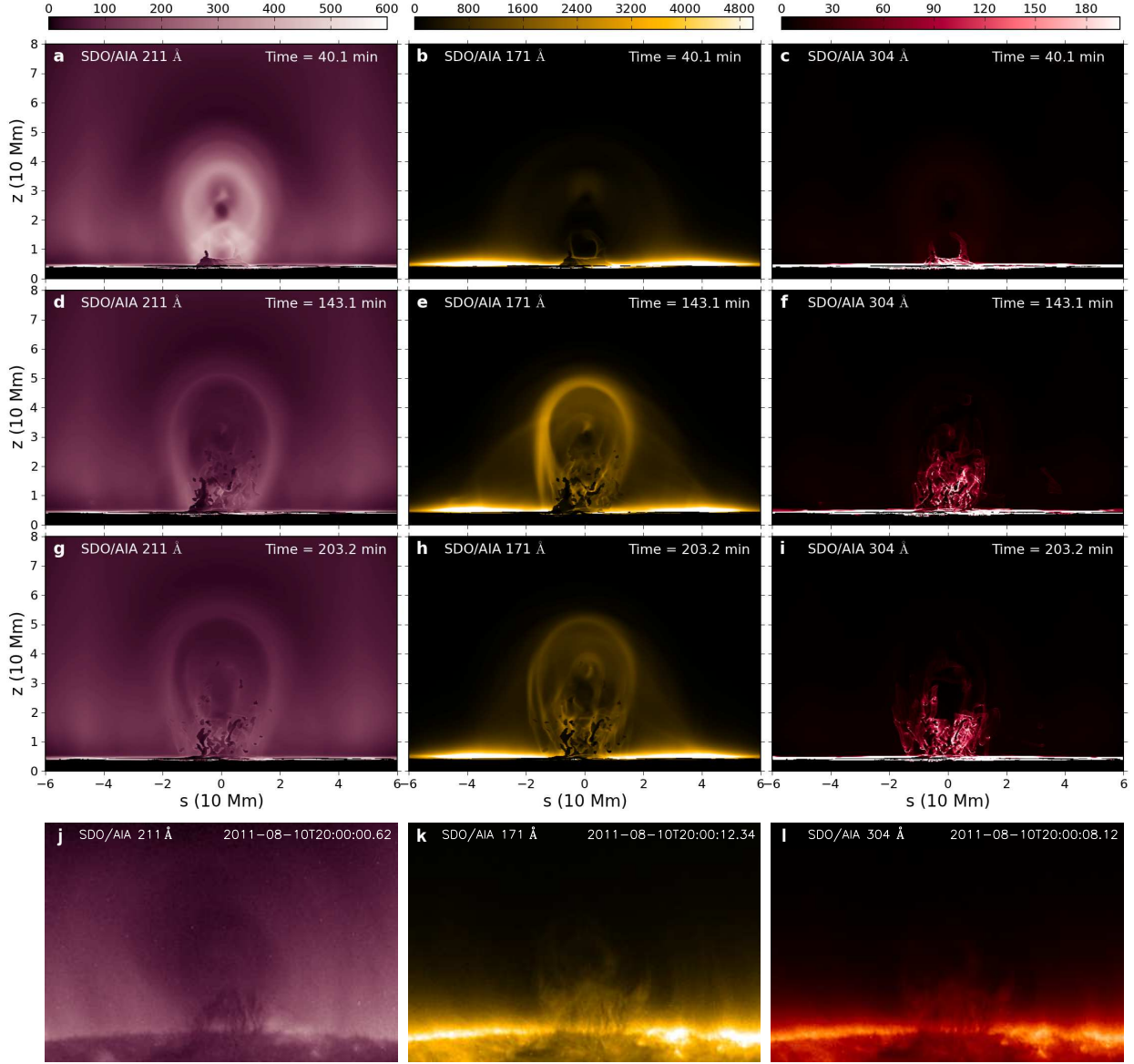}
\caption{Synthetic EUV images of the forming prominence 
viewed along the axis of the prominence and comparison with observations.
Synthetic SDO/AIA EUV images of the simulated prominence at wavebands 211, 171, 
and 304 \AA, at time 40.1 minutes (\textbf{a}--\textbf{c}), 143.1 minutes 
(\textbf{d}--\textbf{f}), and 203.2 minutes (\textbf{g}--\textbf{i}) and 
SDO/AIA observations on a real prominence and its cavity 
(\textbf{j}--\textbf{l}). Supplementary Movie 1 shows the temporal evolution of
these synthetic views.}
\label{fsynax}
\end{figure}

To validate our model, we have made synthetic observations on our simulations with
technique details described in \citet{Xia14},  
for direct comparison with those from the Solar Dynamics Observatory (SDO) 
Atmospheric Imaging Assembly (AIA) instrument. Synthetic EUV images are 
generated from different viewing angles through the AIA wavebands at 211, 193, 
171, and 304 \AA, which have main contribution temperatures around 1.8, 1.5, 
0.8, and 0.08 MK, respectively. To mimic absorption of background extreme 
ultraviolet emission by the prominence plasma, we exclude emission coming 
from behind prominence plasma with density higher than $2\times10^{10}$ cm$^{-3}$. 
To view along the axis of the prominence, we select a horizontal line of sight 
that deviates from the $x$-axis by 8.4 degrees. Viewing along the axis of the 
prominence, the flux rope region becomes brighter and reaches a maximum 
sequentially in the 211, 193, and 171 \AA\, extreme ultraviolet channels at time 55.8 
minutes, 78.7 minutes, and 95.9 minutes, respectively. At time 40.1 minutes, bright
horn-like structures in the 304 \AA\, image protrude from the chromosphere, showing that the
prominence plasma first forms at low altitudes. Later on, dynamic dark clumps and threads, the
dense cores of the condensations, appear in the lower part of the flux rope in the 211 and
171 \AA\, hot channels, while they are bright in the 304 \AA\, cool channel (Figure~\ref{fsynax}).
Smaller clumps tend to form at higher altitudes and fall down into the chromosphere along
curved trajectories. Larger clumps and threads often present vertical pillars collected in
the dipped region of the flux rope. A dark coronal cavity enclosed by a bright elliptical loop
appears above the dark condensations in hot channels, with higher contrast in hotter channels. 
The coronal cavity expands slowly with time and reaches a height of 54 Mm at the end of our 
simulation. In the bottom row of Figure~\ref{fsynax}, we plot AIA observations of a real 
prominence on August 10, 2011 for direct comparison. Our modeled prominence closely resembles 
the real observed one. We prepared a movie (Supplementary Movie 1) to show the temporal 
evolution of the axial view on the modeled prominence in AIA 211, 193, 171, and 304 \AA~ 
channels. While observations on solar prominences are limited to one or two viewing angles,
we can observe our simulated prominence from any viewing angle.
When we put the line of sight along the $y$-axis, we witness the prominence 
formation from the flank. From this vantage point, the weak EUV brightness in the initial 
corona is quite homogeneous until the localized heating induces sudden brightening of 
the flux rope where many thin helical loops prominently appear in the 211 and 193 \AA\, 
images. At time 40.1 minutes, two oblique condensations, far from each other, grow in the 
dipped portions of asymmetric twisted loops. In fact, these widely separated threads are those 
that resemble the horns in the axial view. At time 95.9 minutes, two long condensed threads, 
about 40 Mm long in 304 \AA, stretch horizontally across helical field lines at altitudes 
of about 20 Mm. Then numerous fragmented condensations appear and cluster in several places. 
They follow different curved trajectories of varying shape and size. Many of them slip into 
the left and right footpoints of the flux rope, while others linger in the central dipped 
regions (see Supplementary Movie 2).

\section{CONCLUSIONS AND DISCUSSION}\label{conc}
The model described above demonstrated the physical mechanism responsible 
for the formation of prominences in realistic 3D coronal environments and 
self-consistently explained the vertical flows in thin prominence threads, which
reveal the plasma circulation of long-lived prominences. Their longevity is in analogy 
with the water cycle on Earth where plasma evaporates from the chromosphere 
into the corona, in-situ condensations happen continuously to form threads by 
runaway cooling, while 
prominence threads descend and eventually fall back to the chromosphere, 
dragging along the mass-loaded magnetic field lines. This offers completely new 
insights on how prominence matter gets recycled between chromospheric and coronal 
heights, and revolutionizes all previous models where a more static, large-scale 
structure to model prominences is adopted. Since many prominences erupt into 
interplanetary space at the end of their lives, they are vital ingredients in
coronal mass ejections (CME) which may have severe impact on the terrestrial 
space environment. To better understand prominence eruption and CMEs, 
realistic prominence models are needed. Our model represents a milestone for 
constructing realistic prominence models and can now be used to initiate and
study the eruption phase in details. 

\acknowledgments
C.X. wants to thank FWO (Research Foundation Flanders) for the award of 
postdoctoral fellowship and thank Oliver Porth, Yang Guo, and Junjie Yi for 
discussions. This research was supported by FWO and the Interuniversity 
Attraction Poles Programme by the Belgian Science Policy Office (IAP P7/08 
CHARM). The simulations were conducted on the VSC (Flemish Supercomputer 
Center funded by Hercules foundation and Flemish government) and SuperMUC 
supercomputers (provided by PRACE resources in project pr87di).


\clearpage

\clearpage

\clearpage

\clearpage

\clearpage

\clearpage
\begin{figure}
\includegraphics[width=\textwidth]{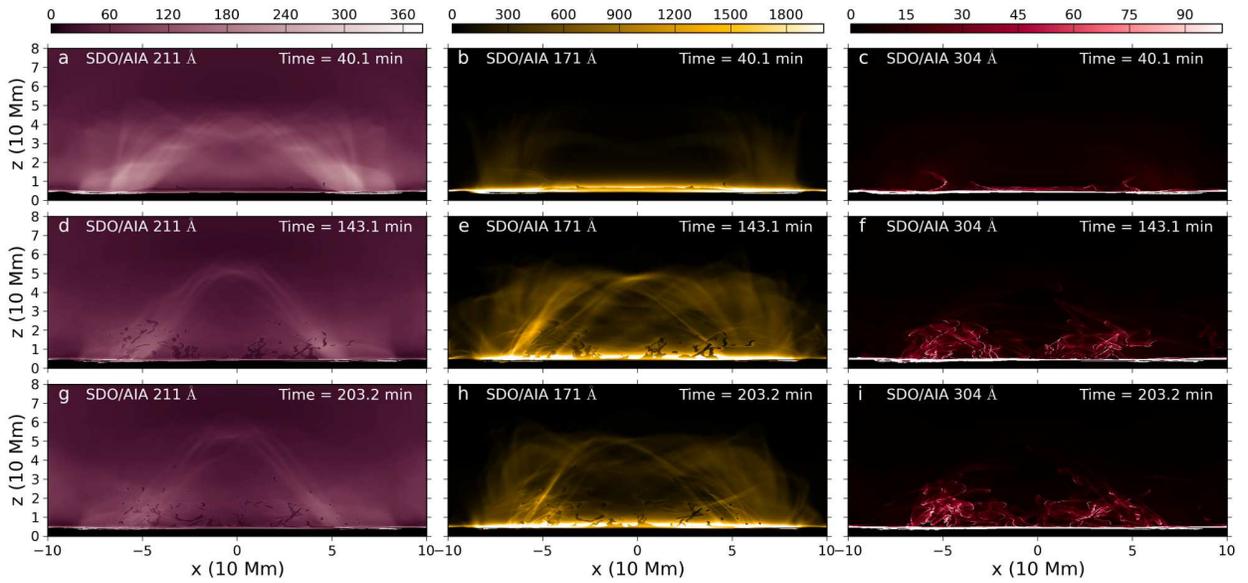}
\caption{Synthetic EUV images of the forming prominence with the line of sight 
along the $y$-axis nearly perpendicular to the axis of the prominence.
Synthetic SDO/AIA EUV images of the simulated prominence at wavebands 211, 171, 
and 304 \AA, at time 40.1 minutes (\textbf{a}--\textbf{c}), 143.1 minutes 
(\textbf{d}--\textbf{f}), and 203.2 minutes (\textbf{g}--\textbf{i}).
Supplementary Movie 2 shows the temporal evolution of these synthetic views.}
\label{fsynfl}
\end{figure}

\end{document}